\pdfoutput=1
\documentclass[12pt,a4paper]{article}
\usepackage[colorinlistoftodos]{todonotes}
\usepackage{ifthen} 
\newboolean{pdflatex}
\setboolean{pdflatex}{true} 
\newboolean{articletitles}
\setboolean{articletitles}{true} 
\newboolean{uprightparticles}
\setboolean{uprightparticles}{false} 
\newboolean{inbibliography}
\setboolean{inbibliography}{false} 
\def\paperauthors{LHCb collaboration} 
\def\paperasciititle{Measurement of CP asymmetries in Lb0->ph- decays} 
\def\papertitle{Measurement of $C\!P$ asymmetries in $\Lb\to ph^{-}$ decays} 
\def\paperkeywords{{High Energy Physics}, {LHCb}} 
\def\papercopyright{\the\year\ CERN for the benefit of the LHCb collaboration} 
\def\paperlicence{CC-BY-4.0 licence}
\def\paperlicenceurl{https://creativecommons.org/licenses/by/4.0/}

\usepackage[top=1in, bottom=1.25in, left=1in, right=1in]{geometry}

\usepackage{microtype}
\usepackage{lineno}  
\usepackage{xspace} 
\usepackage{caption} 

\usepackage{graphicx}  
\usepackage{color}
\usepackage{colortbl}
\graphicspath{{./}} 

\usepackage{amsmath} 
\usepackage{amssymb}
\usepackage{amsfonts}
\usepackage{upgreek} 

\newcommand*\patchAmsMathEnvironmentForLineno[1]{%
    \expandafter\let\csname old#1\expandafter\endcsname\csname #1\endcsname
    \expandafter\let\csname oldend#1\expandafter\endcsname\csname
    end#1\endcsname
    \renewenvironment{#1}%
    {\linenomath\csname old#1\endcsname}%
    {\csname oldend#1\endcsname\endlinenomath}%
}
\newcommand*\patchBothAmsMathEnvironmentsForLineno[1]{%
    \patchAmsMathEnvironmentForLineno{#1}%
    \patchAmsMathEnvironmentForLineno{#1*}%
}
\AtBeginDocument{%
    \patchBothAmsMathEnvironmentsForLineno{equation}%
    \patchBothAmsMathEnvironmentsForLineno{align}%
    \patchBothAmsMathEnvironmentsForLineno{flalign}%
    \patchBothAmsMathEnvironmentsForLineno{alignat}%
    \patchBothAmsMathEnvironmentsForLineno{gather}%
    \patchBothAmsMathEnvironmentsForLineno{multline}%
    \patchBothAmsMathEnvironmentsForLineno{eqnarray}%
}

\usepackage[pdftex,
    pdfauthor={\paperauthors},
    pdftitle={\paperasciititle},
    pdfkeywords={\paperkeywords},
]{hyperref}
\usepackage{hyperxmp}

\usepackage[all]{hypcap} 

\usepackage{xspace}
\usepackage{upgreek}

\def\lhcb   {\mbox{LHCb}\xspace}

\def\babar  {\mbox{BaBar}\xspace}

\def\cdf    {\mbox{CDF}\xspace}

\def\velo   {VELO\xspace}
\def\rich   {RICH\xspace}

\def\MagUp {\mbox{\em Mag\kern -0.05em Up}\xspace}

\ifthenelse{\boolean{uprightparticles}}%
{

    \def\Pmu         {\ensuremath{\upmu}\xspace}
    \def\Pnu         {\ensuremath{\upnu}\xspace}
    
    \def\Ppi         {\ensuremath{\uppi}\xspace}

    \def\Ppsi        {\ensuremath{\uppsi}\xspace}

    \def\PDelta      {\ensuremath{\Delta}\xspace}
    \def\PXi      {\ensuremath{\Xi}\xspace}
    \def\PLambda      {\ensuremath{\Lambda}\xspace}
    \def\PSigma      {\ensuremath{\Sigma}\xspace}
    \def\POmega      {\ensuremath{\Omega}\xspace}
    \def\PUpsilon      {\ensuremath{\Upsilon}\xspace}

    \def\PB      {\ensuremath{\mathrm{B}}\xspace}
    
    \def\PD      {\ensuremath{\mathrm{D}}\xspace}

    \def\PJ      {\ensuremath{\mathrm{J}}\xspace}
    \def\PK      {\ensuremath{\mathrm{K}}\xspace}

    \def\PX      {\ensuremath{\mathrm{X}}\xspace}

    \def\Pb      {\ensuremath{\mathrm{b}}\xspace}
    \def\Pc      {\ensuremath{\mathrm{c}}\xspace}

    \def\Ph      {\ensuremath{\mathrm{h}}\xspace}     
    \def\Pi      {\ensuremath{\mathrm{i}}\xspace}

    \def\Pp      {\ensuremath{\mathrm{p}}\xspace}

    \def\Ps      {\ensuremath{\mathrm{s}}\xspace}

}
{

    \def\Pmu         {\ensuremath{\mu}\xspace}
    \def\Pnu         {\ensuremath{\nu}\xspace}
    
    \def\Ppi         {\ensuremath{\pi}\xspace}

    \def\Ppsi        {\ensuremath{\psi}\xspace}
    
    \mathchardef\PDelta="7101
    \mathchardef\PXi="7104
    \mathchardef\PLambda="7103
    \mathchardef\PSigma="7106
    \mathchardef\POmega="710A
    \mathchardef\PUpsilon="7107
    
    \def\PB      {\ensuremath{B}\xspace}
    
    \def\PD      {\ensuremath{D}\xspace}

    \def\PJ      {\ensuremath{J}\xspace}
    \def\PK      {\ensuremath{K}\xspace}

    \def\PX      {\ensuremath{X}\xspace}

    \def\Pb      {\ensuremath{b}\xspace}
    \def\Pc      {\ensuremath{c}\xspace}

    \def\Ph      {\ensuremath{h}\xspace}              
    \def\Pi      {\ensuremath{i}\xspace}

    \def\Pp      {\ensuremath{p}\xspace}

    \def\Ps      {\ensuremath{s}\xspace}

}

\makeatletter
\ifcase \@ptsize \relax
    \newcommand{\miniscule}{\@setfontsize\miniscule{4}{5}}
\or
    \newcommand{\miniscule}{\@setfontsize\miniscule{5}{6}}
\or
    \newcommand{\miniscule}{\@setfontsize\miniscule{5}{6}}
\fi
\makeatother

\DeclareRobustCommand{\optbar}[1]{\shortstack{{\miniscule (\rule[.5ex]{1.25em}{.18mm})}
\\ [-.7ex] $#1$}}

\def\mun        {{\ensuremath{\Pmu^-}}\xspace} 

\def\mumu       {{\ensuremath{\Pmu^+\Pmu^-}}\xspace}

\def\neub       {{\ensuremath{\overline{\Pnu}}}\xspace}

\def\neumb      {{\ensuremath{\neub_\mu}}\xspace}

\def\squark    {{\ensuremath{\Ps}}\xspace}

\def\cquark    {{\ensuremath{\Pc}}\xspace}

\def\bquark    {{\ensuremath{\Pb}}\xspace}

\def\pion   {{\ensuremath{\Ppi}}\xspace}
\def\piz    {{\ensuremath{\pion^0}}\xspace}
\def\pip    {{\ensuremath{\pion^+}}\xspace}
\def\pim    {{\ensuremath{\pion^-}}\xspace}

\def\kaon    {{\ensuremath{\PK}}\xspace}
\def\Kbar    {{\kern 0.2em\overline{\kern -0.2em \PK}{}}\xspace}

\def\KorKbar    {\kern 0.18em\optbar{\kern -0.18em K}{}\xspace}

\def\Kp      {{\ensuremath{\kaon^+}}\xspace}
\def\Km      {{\ensuremath{\kaon^-}}\xspace}

\def\KS      {{\ensuremath{\kaon^0_{\mathrm{ \scriptscriptstyle S}}}}\xspace}

\def\Dbar    {{\kern 0.2em\overline{\kern -0.2em \PD}{}}\xspace}
\def\D       {{\ensuremath{\PD}}\xspace}

\def\DorDbar    {\kern 0.18em\optbar{\kern -0.18em D}{}\xspace}
\def\Dz      {{\ensuremath{\D^0}}\xspace}
\def\Dzb     {{\ensuremath{\Dbar{}^0}}\xspace}
\def\Dp      {{\ensuremath{\D^+}}\xspace}

\def\Dstarp  {{\ensuremath{\D^{*+}}}\xspace}

\def\B       {{\ensuremath{\PB}}\xspace}
\def\Bbar    {{\ensuremath{\kern 0.18em\overline{\kern -0.18em \PB}{}}}\xspace}

\def\BorBbar    {\kern 0.18em\optbar{\kern -0.18em B}{}\xspace}
\def\Bz      {{\ensuremath{\B^0}}\xspace}

\def\Bu      {{\ensuremath{\B^+}}\xspace}

\def\Bp      {{\ensuremath{\Bu}}\xspace}

\def\Bd      {{\ensuremath{\B^0}}\xspace}
\def\Bs      {{\ensuremath{\B^0_\squark}}\xspace}
\def\Bsb     {{\ensuremath{\Bbar{}^0_\squark}}\xspace}

\def\Bds     {{\ensuremath{\B_{(\squark)}^0}}\xspace}

\def\BdorBs  {\Bds}

\def\jpsi     {{\ensuremath{{\PJ\mskip -3mu/\mskip -2mu\Ppsi\mskip 2mu}}}\xspace}

\def\Y#1S{\ensuremath{\PUpsilon{(#1S)}}\xspace}

\def\proton      {{\ensuremath{\Pp}}\xspace}
\def\antiproton  {{\ensuremath{\overline \proton}}\xspace}

\def\Xires       {{\ensuremath{\PXi}}\xspace}

\def\Lz          {{\ensuremath{\PLambda}}\xspace}
\def\Lbar        {{\ensuremath{\kern 0.1em\overline{\kern -0.1em\PLambda}}}\xspace}
\def\LorLbar    {\kern 0.18em\optbar{\kern -0.18em \PLambda}{}\xspace}

\def\Lb      {{\ensuremath{\Lz^0_\bquark}}\xspace}
\def\Lbbar   {{\ensuremath{\Lbar{}^0_\bquark}}\xspace}
\def\Lc      {{\ensuremath{\Lz^+_\cquark}}\xspace}

\def\Xibz    {{\ensuremath{\Xires^0_\bquark}}\xspace}
\def\Xibm    {{\ensuremath{\Xires^-_\bquark}}\xspace}

\def\Xicp    {{\ensuremath{\Xires^+_\cquark}}\xspace}

\newcommand{\decay}[2]{\ensuremath{#1\!\to #2}\xspace}         

\def\to                 {\ensuremath{\rightarrow}\xspace}

\def\CP                {{\ensuremath{C\!P}}\xspace}

\def\BdTopipi     {\decay{\Bd}{\pip\pim}}
\def\BdToKpi      {\decay{\Bd}{\Kp\pim}}
\def\BsToKK       {\decay{\Bs}{\Kp\Km}}

\def\LbTopK       {\decay{\Lb}{\proton\Km}}

\def\LbToppi      {\decay{\Lb}{\proton\pim}}

\def\LbToph       {\decay{\Lb}{\proton h^{-}}}
\def\LbToLcpi     {\decay{\Lb}{\Lc\pim}}

\def\AT#1     {\ensuremath{A_{\mathrm{T}}^{#1}}\xspace}           

\def\C#1      {\ensuremath{\mathcal{C}_{#1}}\xspace}                       
\def\Cp#1     {\ensuremath{\mathcal{C}_{#1}^{'}}\xspace}                    
\def\Ceff#1   {\ensuremath{\mathcal{C}_{#1}^{\mathrm{(eff)}}}\xspace}        
\def\Cpeff#1  {\ensuremath{\mathcal{C}_{#1}^{'\mathrm{(eff)}}}\xspace}       
\def\Ope#1    {\ensuremath{\mathcal{O}_{#1}}\xspace}                       
\def\Opep#1   {\ensuremath{\mathcal{O}_{#1}^{'}}\xspace}                    

\newcommand{\tev}{\ifthenelse{\boolean{inbibliography}}{\ensuremath{~T\kern -0.05em eV}}{\ensuremath{\mathrm{\,Te\kern -0.1em V}}}\xspace}
\newcommand{\gev}{\ensuremath{\mathrm{\,Ge\kern -0.1em V}}\xspace}
\newcommand{\mev}{\ensuremath{\mathrm{\,Me\kern -0.1em V}}\xspace}
\newcommand{\kev}{\ensuremath{\mathrm{\,ke\kern -0.1em V}}\xspace}
\newcommand{\ev}{\ensuremath{\mathrm{\,e\kern -0.1em V}}\xspace}
\newcommand{\gevc}{\ensuremath{{\mathrm{\,Ge\kern -0.1em V\!/}c}}\xspace}
\newcommand{\mevc}{\ensuremath{{\mathrm{\,Me\kern -0.1em V\!/}c}}\xspace}
\newcommand{\gevcc}{\ensuremath{{\mathrm{\,Ge\kern -0.1em V\!/}c^2}}\xspace}
\newcommand{\gevgevcccc}{\ensuremath{{\mathrm{\,Ge\kern -0.1em V^2\!/}c^4}}\xspace}
\newcommand{\mevcc}{\ensuremath{{\mathrm{\,Me\kern -0.1em V\!/}c^2}}\xspace}

\def\m    {\ensuremath{\mathrm{ \,m}}\xspace}

\def\mum  {\ensuremath{{\,\upmu\mathrm{m}}}\xspace}

\def\invfb   {\ensuremath{\mbox{\,fb}^{-1}}\xspace}

\newcommand{\chisqip}{\ensuremath{\chi^2_{\text{IP}}}\xspace}

\def\gsim{{~\raise.15em\hbox{$>$}\kern-.85em
            \lower.35em\hbox{$\sim$}~}\xspace}
\def\lsim{{~\raise.15em\hbox{$<$}\kern-.85em
            \lower.35em\hbox{$\sim$}~}\xspace}

\def\pt         {\ensuremath{p_{\mathrm{ T}}}\xspace}
\def\ptot       {\ensuremath{p}\xspace}

\def\evtgen     {\mbox{\textsc{EvtGen}}\xspace}

\def\geant      {\mbox{\textsc{Geant4}}\xspace}

\def\photos     {\mbox{\textsc{Photos}}\xspace}

\def\pythia     {\mbox{\textsc{Pythia}}\xspace}

\def\tell1  {TELL1\xspace}
\def\ukl1   {UKL1\xspace}

\newcommand{\eg}{\mbox{\itshape e.g.}\xspace}
\newcommand{\ie}{\mbox{\itshape i.e.}\xspace}

\def\hp    {{\ensuremath{\Ph^+}}\xspace}

\def\hm    {{\ensuremath{\Ph^-}}\xspace}

\def\APLb {\ensuremath{A_{\mathrm P}^\Lb}\xspace}

\def\ADhm {\ensuremath{A_{\mathrm D}^{\hm}}\xspace}
\def\ADhp {\ensuremath{A_{\mathrm D}^{\hp}}\xspace}
\def\ADKm {\ensuremath{A_{\mathrm D}^{\Km}}\xspace}

\def\ADpip {\ensuremath{A_{\mathrm D}^{\pip}}\xspace}
\def\ADpim {\ensuremath{A_{\mathrm D}^{\pim}}\xspace}
\def\ADp {\ensuremath{A_{\mathrm D}^\proton}\xspace}

\def\Araw {\ensuremath{A_{\mathrm{raw}}}\xspace}

\def\ACPph {\ensuremath{A_\CP^{\proton \hm}}\xspace}
\def\ACPpK {\ensuremath{A_\CP^{\proton \Km}}\xspace}
\def\ACPppi {\ensuremath{A_\CP^{\proton \pim}}\xspace}
\def\ACPLcpi {\ensuremath{A_\CP^{\Lc \pim}}\xspace}
\def\Arawph {\ensuremath{A_\mathrm{raw}^{\proton \hm}}\xspace}
\def\ArawpK {\ensuremath{A_\mathrm{raw}^{\proton \Km}}\xspace}
\def\Arawppi {\ensuremath{A_\mathrm{raw}^{\proton \pim}}\xspace}
\def\APID {\ensuremath{A_\mathrm{PID}}\xspace}
\def\APIDph {\ensuremath{A_\mathrm{PID}^{\proton \hm}}\xspace}

\def\ATph  {\ensuremath{A_\mathrm{T}^{\proton \hm}}\xspace}

\def\SpK  {\ensuremath{S_{\proton \Km}}\xspace}
\def\Sppi  {\ensuremath{S_{\proton \pim}}\xspace}

\newcommand{\lhcborcid}[1]{\href{https://orcid.org/#1}{\hspace*{0.1em}\raisebox{-0.45ex}{\includegraphics[width=1em]{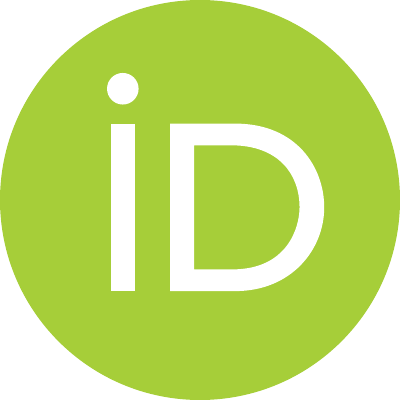}}}}

\usepackage{cite} 
\usepackage{mciteplus}
\usepackage{multirow}
\usepackage{csquotes}

\usepackage{tabularx}
\newcolumntype{L}[1]{>{\raggedright\arraybackslash}p{#1}}
\newcolumntype{C}[1]{>{\centering\arraybackslash}p{#1}}
\newcolumntype{R}[1]{>{\raggedleft\arraybackslash}p{#1}}

\RequirePackage[normalem]{ulem} 
\RequirePackage{color}\definecolor{RED}{rgb}{1,0,0}\definecolor{BLUE}{rgb}{0,0,1} 

\usepackage{xfp}
\newcommand{\rounded}[2][1]{\num[round-mode=places,round-precision=#1]{#2}}

\def\ArawpkRunOneError{1.549}

\def\ArawppiRunOneError{1.856}

\def\ArawpkRunTwoError{0.745}

\def\ArawppiRunTwoError{0.93}

\def\ACPcorrRunOne{2.7}
\def\ACPcorrRunTwo{9.8}
\def\ACPcorrRunOneTwo{9.6}

\def\ACPpkRunOneCentral{-0.27}
\def\ACPpkRunOneError{1.65}
\def\ACPpkRunOneStat{\ArawpkRunOneError}
\def\ACPpkRunOneStatW{\fpeval{1/\ACPpkRunOneStat/\ACPpkRunOneStat}}
\def\ACPpkRunOneSyst{\fpeval{sqrt(\ACPpkRunOneError*\ACPpkRunOneError - \ACPpkRunOneStat*\ACPpkRunOneStat)}}

\def\ACPpkRunTwoCentral{-1.39}
\def\ACPpkRunTwoError{0.85}
\def\ACPpkRunTwoStat{\ArawpkRunTwoError}
\def\ACPpkRunTwoStatW{\fpeval{1/\ACPpkRunTwoStat/\ACPpkRunTwoStat}}
\def\ACPpkRunTwoSyst{\fpeval{sqrt(\ACPpkRunTwoError*\ACPpkRunTwoError - \ACPpkRunTwoStat*\ACPpkRunTwoStat)}}

\def\ACPpkCentral{-1.14}
\def\ACPpkError{0.76}
\def\ACPpkStat{\fpeval{1/sqrt(\ACPpkRunOneStatW + \ACPpkRunTwoStatW)}}
\def\ACPpkSyst{\fpeval{sqrt(\ACPpkError*\ACPpkError - \ACPpkStat*\ACPpkStat)}}

\def\ACPppiRunOneCentral{-0.59}
\def\ACPppiRunOneError{1.93}
\def\ACPppiRunOneStat{\ArawppiRunOneError}
\def\ACPppiRunOneStatW{\fpeval{1/\ACPppiRunOneStat/\ACPppiRunOneStat}}
\def\ACPppiRunOneSyst{\fpeval{sqrt(\ACPppiRunOneError*\ACPppiRunOneError - \ACPppiRunOneStat*\ACPppiRunOneStat)}}

\def\ACPppiRunTwoCentral{0.42}
\def\ACPppiRunTwoError{1.02}
\def\ACPppiRunTwoStat{\ArawppiRunTwoError}
\def\ACPppiRunTwoStatW{\fpeval{1/\ACPppiRunTwoStat/\ACPppiRunTwoStat}}
\def\ACPppiRunTwoSyst{\fpeval{sqrt(\ACPppiRunTwoError*\ACPppiRunTwoError - \ACPppiRunTwoStat*\ACPppiRunTwoStat)}}

\def\ACPppiCentral{0.20}
\def\ACPppiError{0.91}
\def\ACPppiStat{\fpeval{1/sqrt(\ACPppiRunOneStatW + \ACPppiRunTwoStatW)}}
\def\ACPppiSyst{\fpeval{sqrt(\ACPppiError*\ACPppiError - \ACPppiStat*\ACPppiStat)}}

\usepackage{siunitx}
\usepackage{cleveref}
\usepackage{booktabs}
\usepackage{longtable} 
\begin{document}
\renewcommand{\thefootnote}{\fnsymbol{footnote}}
\setcounter{footnote}{1}

\begin{titlepage}
    \pagenumbering{roman}

    \vspace*{-1.5cm}
    \centerline{\large EUROPEAN ORGANIZATION FOR NUCLEAR RESEARCH (CERN)}
    \vspace*{1.5cm}
    \noindent
    \begin{tabular*}{\linewidth}{lc@{\extracolsep{\fill}}r@{\extracolsep{0pt}}}
        \\
        & & CERN-EP-2024-330 \\  
        & & LHCb-PAPER-2024-048 \\  
        & & May 8, 2025 \\
        & & \\
    \end{tabular*}

    \vspace*{4.0cm}

    {\normalfont\bfseries\boldmath\huge
        \begin{center}
            \papertitle
        \end{center}
    }

    \vspace*{2.0cm}

    \begin{center}
        \paperauthors\footnote{Authors are listed at the end of this paper.}
    \end{center}

    \vspace{\fill}

    \begin{abstract}
        \noindent
        A search for $C\!P$ violation in $\Lb\rightarrow pK^-$ and $\Lb\rightarrow p\pi^-$ decays is presented using the full Run~1 and Run~2 data samples of $pp$ collisions collected with the LHCb detector, corresponding to an integrated luminosity of 9\invfb at center-of-mass energies of 7, 8, and 13\tev.
        For the Run~2 data sample, the $C\!P$-violating asymmetries  are measured to be $A_{C\!P}^{pK^-} = (-1.4 \pm 0.7 \pm 0.4)\%$ and $A_{C\!P}^{p\pi^-} = (0.4 \pm 0.9 \pm 0.4)\%$, where the first uncertainty is statistical and the second is systematic.
        Following significant improvements in the evaluation of systematic uncertainties compared to the previous \lhcb measurement, the Run~1 dataset is reanalyzed to update the corresponding results.
        When combining the Run~2 and updated Run~1 measurements, the final results are found to be $A_{C\!P}^{pK^-} = (-1.1 \pm 0.7 \pm 0.4)\%$ and $A_{C\!P}^{p\pi^-} = (0.2 \pm 0.8 \pm 0.4)\%$, constituting the most precise measurements of these asymmetries to date.
    \end{abstract}

    \vspace*{2.0cm}

    \begin{center}
        Published in Phys.~Rev.~D 111 (2025) 092004
    \end{center}

    \vspace{\fill}

    {\footnotesize
        \centerline{\copyright~\papercopyright. \href{\paperlicenceurl}{\paperlicence}.}}
    \vspace*{2mm}

\end{titlepage}


\newpage
\setcounter{page}{2}
\mbox{~}

\cleardoublepage

\renewcommand{\thefootnote}{\arabic{footnote}}
\setcounter{footnote}{0}
\pagestyle{plain} 
\setcounter{page}{1}
\pagenumbering{arabic}
\section{Introduction}
\label{sec:Introduction}
\noindent
The Standard Model (SM) of particle physics predicts the noninvariance of weak interactions under the combined application of charge conjugation ($C$) and parity ($P$) transformations via the Cabibbo--Kobayashi--Maskawa mechanism~\cite{Cabibbo:1963yz,Kobayashi:1973fv}.
In the baryon sector, many searches for \CP violation (CPV) have been performed~\cite{LHCb-PAPER-2013-061,LHCb-PAPER-2014-020,LHCb-PAPER-2016-002,LHCb-PAPER-2016-004,LHCb-PAPER-2016-030,LHCb-PAPER-2019-028,LHCb-PAPER-2016-059,LHCb-PAPER-2017-044,LHCb-PAPER-2018-001,CDF:2011ubb,CDF:2014pzb,LHCb-PAPER-2018-025,LHCb-PAPER-2018-044,LHCb-PAPER-2019-026,LHCb-PAPER-2020-017,LHCb-PAPER-2021-027,LHCb-PAPER-2021-030}, with the only evidence coming from a recent analysis of the decay \decay{\Lb}{\Lz\Kp\Km}~\cite{LHCb-PAPER-2024-043}.
In this scenario, \LbTopK and \LbToppi decays\footnote{Unless stated otherwise, the inclusion of charge-conjugate processes is implied throughout this paper.} are promising candidates to search for CPV, as they are mediated by the same quark-level transitions that contribute to charmless two-body \decay{\Bz}{\Kp\pim} and \decay{\Bs}{\pip\Km} decays, where CPV is well established~\cite{Belle:2008alz,LHCb-PAPER-2011-029,BellePhysRevD.87.031103,BaBarPhysRevD.87.052009,CDF:2014pzb,LHCb-PAPER-2018-006,LHCb-PAPER-2020-029}.

In the SM, predictions for the \CP asymmetries in decays of the \Lb baryon to two-body charmless final states $\proton \Km$ or $\proton \pim$ are characterized by relatively large uncertainties. These arise from difficulties in computing the hadronic parameters due to residual quantum chromodynamic (QCD) interactions between the quarks, with values of a few percent reported both in the generalized factorization approach~\cite{Hsiao:2017tif,Zhu:2018jet} and in the MIT bag model~\cite{Geng:2020ofy}.
Recently, it has been suggested that the destructive interference between different partial waves in these two decays could explain the smallness of the measured \CP asymmetries~\cite{Yu:2024cjd}.
Measurements of these quantities were performed by the \cdf and \lhcb collaborations~\cite{CDF:2011ubb,CDF:2014pzb,LHCb-PAPER-2018-025}.
The world averages, dominated by the Run~1 \lhcb measurement, show asymmetries compatible with zero within uncertainties of $2.2$\% and $2.9\%$ for \LbTopK and \LbToppi decays, respectively~\cite{PDG2024}.

This paper presents a search for \CP violation in \LbTopK and \LbToppi decays, using proton-proton ($pp$) collision data collected with the LHCb detector at center-of-mass energies of 7, 8, and 13\tev and corresponding to 9\invfb of integrated luminosity.
The dataset has been collected in two data-taking campaigns, called Run~1 (2011--2012) and Run~2 (2015--2018).
The \CP asymmetry is defined as
\begin{equation}
    A_\CP^f \equiv \frac{\Gamma(\Lb \to f) - \Gamma(\Lbbar \to \overline{f})}{\Gamma(\Lb \to f) + \Gamma(\Lbbar \to \overline{f})},
\end{equation}
where $\Gamma$ is the partial width of the given decay, with $f \in \{\proton\Km,\proton \pim\}$ and $\overline{f} \in \{\antiproton\Kp,\antiproton \pip\}$.

The \textcolor{black}{remainder} of the paper is organized as follows.
After a brief description of the detector, trigger and simulation in Sec.~\ref{sec:DetectorTriggerAndSimulation}, the formalism needed to relate the physical \CP asymmetries to the experimental measurements is presented in Sec.~\ref{sec:Formalism}.
The event selection and fits to the invariant-mass distributions are described in Secs.~\ref{sec:EventSelection} and~\ref{sec:InvariantMassFit}, respectively.
The determination of instrumental asymmetries is discussed in Sec.~\ref{sec:InstrumentalAsymmetries}, while the estimation of systematic uncertainties are presented in Sec.~\ref{sec:SystematicUncertaintiesAndCombination}.
Finally, the results are given and conclusions drawn in Sec.~\ref{sec:ResultsAndConclusions}.

\section{Detector, trigger and simulation}
\label{sec:DetectorTriggerAndSimulation}
The \lhcb detector~\cite{LHCb-DP-2008-001,LHCb-DP-2014-002} is a single-arm forward spectrometer covering the pseudorapidity range $2<\eta <5$, designed for the study of particles containing \bquark or \cquark quarks.
The detector includes a high-precision tracking system consisting of a silicon-strip vertex detector (\velo) surrounding the $pp$ interaction region, a large-area silicon-strip detector located upstream of a dipole magnet with a bending power of about \SI{4}{\tesla\m}, and three stations of silicon-strip detectors and straw drift tubes placed downstream of the magnet.
The tracking system provides a measurement of the momentum, \ptot, of charged particles with a relative uncertainty that varies from 0.5\% at low momentum to 1.0\% at 200\gevc.
The minimum distance of a track to a primary vertex (PV), the impact parameter (IP), is measured with a resolution of $(15+29/\pt)\mum$, where \pt is the component of the momentum transverse to the beam, in \gevc.
The magnetic field deflects particles in opposite
directions based on their charge, which can lead to detection asymmetries.
To reduce such effects, the magnet polarity is reversed periodically throughout the data taking.
Different types of charged hadrons are distinguished using information from two ring-imaging Cherenkov (\rich) detectors~\cite{LHCb-DP-2012-003}.
Photons, electrons and hadrons are identified by a calorimeter system consisting of a preshower and a scintillating-pad detector (SPD), an electromagnetic and a hadronic calorimeter. Muons are identified by a system composed of alternating layers of iron and multiwire proportional chambers.

The online event selection is performed by a trigger, which consists of a hardware stage followed by a two-level software stage which applies a full event reconstruction.
At the hardware trigger stage, events are required to contain a hadron with transverse energy deposited in the calorimeters greater than approximately 3.7\gev in Run~1 and 4.0\gev in Run~2.
The first stage of the software trigger requires the presence of at least one charged particle with $\pt > 1.6\gevc$ and consistent with originating from a displaced secondary vertex.
In the second software stage, a specific algorithm is used which inclusively reconstructs decays of beauty hadrons to charged two-body final states.
During Run~2, in between the two software stages, an alignment and calibration of the detector were performed in near real-time~\cite{LHCb-PROC-2015-011}, ensuring the availability of high-quality tracking and particle-identification (PID) information at the trigger level.

Simulation is used to optimize the offline event selection and to develop the invariant-mass models to be applied to data.
In the simulation, $pp$ collisions are generated using \pythia~\cite{Sjostrand:2007gs} with a specific \lhcb configuration~\cite{LHCb-PROC-2010-056}.
Decays of unstable particles are described by \evtgen~\cite{Lange:2001uf}, in which final-state radiation is generated using \photos~\cite{Golonka:2005pn}.
The interaction of the generated particles with the detector, and its response, are implemented using the \geant toolkit~\cite{Allison:2006ve, *Agostinelli:2002hh} as described in Ref.~\cite{LHCb-PROC-2011-006}.

\section{Analysis strategy}
\label{sec:Formalism}
Two different strategies have been employed to extract the \CP asymmetries from the Run~1 and Run~2 samples, due to the absence of a measurement of the \Lb baryon production asymmetry for the latter.
For the Run~1 samples, the \CP asymmetries of \LbTopK and \LbToppi decays are computed as the sums of various experimental quantities
\begin{align}
    \ACPph & = \Arawph - \ADp - \ADhm - \APIDph - \ATph - \APLb, \label{eq:ACPph}
\end{align}
where $\Arawph$ is the measured raw asymmetry between the yields of the decays \decay{\Lb}{\proton\hm} and \decay{\Lbbar}{\antiproton\hp}, with $h\in\{K,\,\pi\}$; \ADp and \ADhm are the asymmetries between the detection efficiencies of a given hadron and its charge conjugate; $\APIDph$ is the asymmetry between the PID efficiencies for the final states \proton\hm and \antiproton\hp; \ATph is the asymmetry between the trigger efficiencies for the particles in the final states \proton\hm and \antiproton\hp; and \APLb is the asymmetry between the production cross-sections of \Lb and \Lbbar baryons.
This linear approximation has good accuracy due to the smallness of the terms involved.

The raw asymmetry is defined as
\begin{equation}
    \Araw^f \equiv \frac{N(\Lb \to f) - N(\Lbbar \to \overline{f})}{N(\Lb \to f) + N(\Lbbar \to \overline{f})}, \label{eq:Araw}
\end{equation}
where $N$ denotes the observed signal yield for the given decay, obtained in this analysis by means of extended binned maximum-likelihood fits to the invariant-mass distributions of the 8 possible final states, namely \proton\Km, \antiproton\Kp, \proton\pim, \antiproton\pip, \Kp\pim, \Km\pip, \Kp\Km, and \pip\pim, where the last four are used to constrain the yields of misidentified backgrounds in the signal channels.

The detection asymmetries are defined as
\begin{equation}
    \label{eq:DetectionAsymmetries}
    \ADhp \equiv \frac{\varepsilon_\mathrm{rec}^{\hp}-\varepsilon_\mathrm{rec}^{\hm}}{\varepsilon_\mathrm{rec}^{\hp}+\varepsilon_\mathrm{rec}^{\hm}}\textcolor{black}{\equiv -\ADhm,}
\end{equation}
where $\varepsilon_\mathrm{rec}^{h^{\pm}}$, with $h\in\{\proton,\kaon,\pi\}$, is the total efficiency to reconstruct the given charged particle, excluding PID and trigger requirements.
Such asymmetries are mostly due to the different interaction cross-sections of particles and antiparticles with the detector material, \textcolor{black}{and by construction are asymmetric with under charge reversal}.
The pion detection asymmetry has been measured from the ratio of fully to partially reconstructed \decay{\Dstarp}{(\decay{\Dz}{\Km\pip\pip\pim})\pip} decays~\cite{LHCb-PAPER-2016-013}.
The kaon detection asymmetry is obtained by first measuring the detection asymmetry of the \Kp\pim pair via the subtraction of the raw asymmetries between the \decay{\Dp}{\KS\pip} and \decay{\Dp}{\Km\pip\pip} decay modes, and then removing the single pion detection asymmetry.
The effects of \CP violation and detection asymmetries for the neutral kaons are corrected according to the strategy used in Ref.~\cite{LHCb-PAPER-2014-013}.
The proton detection asymmetry is measured following the procedure described in Ref.~\cite{LHCB-PAPER-2021-016}, using as input the flight distance of protons traversing the detector in units of the nuclear interaction length.
The method is validated with a combined use of a detailed \lhcb detector simulation, external measurements of cross-sections of proton and antiproton scattering on deuterium targets, and information from a large calibration sample of \decay{\Lz}{\proton\pim} decays.
The measurement is performed with Run~2 data and used for both Run~1 and Run~2 samples.
The validity of this approach is based on the assumption that the material budget of the detector did not change significantly between the two data-taking periods, as the only change to the detector was the removal of the aerogel radiator from one of the two RICH detectors.

The PID asymmetries are measured from large calibration samples of \mbox{\decay{\Dstarp}{(\decay{\Dz}{\Km\pip})\pip}} decays for pions and kaons, and \decay{\Lz}{\proton\pim} and \decay{\Lc}{\proton\Km\pip} decays for protons\cite{LHCb-DP-2018-001}.
They are defined as
\begin{equation}
    \label{eq:PIDAsym}
    \APIDph \equiv \frac{\varepsilon_\mathrm{PID}^{\proton\hm}-\varepsilon_\mathrm{PID}^{\antiproton\hp}}{\varepsilon_\mathrm{PID}^{\proton\hm}+\varepsilon_\mathrm{PID}^{\antiproton\hp}},
\end{equation}
where $\varepsilon_\mathrm{PID}^{f}$ is the PID efficiency for a final state $f$ given a set of PID requirements.

Asymmetries arising from different efficiencies on oppositely charged particles for the hardware and software trigger, \ATph, may bias the results. They are defined similarly to Eq.~\eqref{eq:PIDAsym}, and are estimated with data-driven techniques, whose details are given in Sec.~\ref{sec:InstrumentalAsymmetries}.

The \Lb production asymmetry is defined as
\begin{equation}
    \APLb \equiv \frac{\sigma(\Lb) - \sigma(\Lbbar)}{\sigma(\Lb) + \sigma(\Lbbar)},
\end{equation}
where $\sigma$ denotes the inclusive production cross-section in the LHCb acceptance.
The Run~1 production asymmetry has been measured using semileptonic \decay{\Lb}{\Lc\mun\neumb\PX} decays in Ref.~\cite{LHCB-PAPER-2021-016}.

For the Run~2 sample, there is no available measurement of \APLb, therefore a different strategy to obtain the \CP asymmetries has been devised.
Employing a control sample of \decay{\Lb}{(\decay{\Lc}{\proton\Km\pip})\pim} decays, the following formulae are used:
\begin{align}
    \ACPpK  & = \Delta\Araw - \Delta\ADp - \Delta\ADKm - \Delta\APID - \Delta\APLb - \Delta\AT{} - \ADpim -  \ADpip + \ACPLcpi, \label{eq:ACPpKRun2}  \\
    \ACPppi & = \Delta\Araw - \Delta\ADp - \Delta\ADpim - \Delta\APID - \Delta\APLb - \Delta\AT{} - \ADKm -  \ADpip + \ACPLcpi, \label{eq:ACPppiRun2}
\end{align}
where $\Delta A$ indicates the difference between the asymmetries determined for the signal and control modes, and \ADpim, \ADpip, and \ADKm are the detection asymmetries of particles from the control sample that are not accounted for in $\Delta\ADhm$.
The \CP asymmetry in the control sample, \ACPLcpi, is expected to be zero in the SM, within the experimental uncertainties of this analysis.
The control sample is weighted to equalize the two-dimensional distribution in the $(\pt, \eta)$ plane of the \Lb baryon with that from \LbToph decays, such that the $\Delta\APLb{}$ term is zero by construction in Eqs.~\eqref{eq:ACPpKRun2} and \eqref{eq:ACPppiRun2}.

Apart from this change in strategy, there is also a different method used to determine the pion detection asymmetry for Run~2.
It is measured as the ratio of pion yields from \decay{\KS}{\pip\pim} decays reconstructed with all the tracking detectors to those reconstructed using only the track segment in the \velo~\cite{LHCb-PAPER-2021-009}.

\section{Event selection}
\label{sec:EventSelection}

An optimization of the offline event selection is performed for the Run~2 sample, using the same strategy developed for the previous Run~1 analysis.
To start, \bquark-hadron candidates are reconstructed in the second stage of the software trigger, combining pairs of oppositely charged tracks with $\pt > 1\gevc$, that are inconsistent with originating from any PV and are required to form a common vertex.
The candidates must have an invariant mass, computed assigning the pion hypothesis to both final-state tracks, in the range between 4.6 and 6.4\gevcc for Run~1 and between 4.7 and 6.2\gevcc for Run~2.
Finally, each \bquark-hadron candidate is required to be consistent with originating from a PV.
\textcolor{black}{In addition to the event selection strategy}, fiducial requirements in the $(p,\eta)$ plane are applied to both tracks in the final state, to ensure kinematic coverage of the samples used to calibrate the PID performances.
The final-state particles passing the fiducial requirements have their momentum (in \gevc) and pseudorapidity lying inside the polygon of vertices \mbox{($p$,$\eta$) $\in$ \{(0, 2)}, (25, 2), (120, 3.75), (123, 4.2), (0, 4.2)\}.

Particle-identification criteria are used to divide the data sample into mutually exclusive subsamples corresponding to the final-state hypotheses $\proton \Km$, $\antiproton \Kp$, $\proton \pim$, $\antiproton \pip$, $\Kp\pim$, $\Km\pip$, $\Kp\Km$ and $\pip\pim$.
The event selection also includes requirements on the output of a boosted decision tree (BDT) classifier~\cite{TMVA4}, used to reject candidates formed with random combinations of oppositely charged tracks (combinatorial background).
The BDT combines the information from the following variables: the \textcolor{black}{smallest} and \textcolor{black}{largest} transverse momentum of the \bquark-hadron decay products; the \textcolor{black}{smallest} and \textcolor{black}{largest} of their impact parameters; the \textcolor{black}{smallest} and \textcolor{black}{largest} of the logarithms of their \chisqip values, where \chisqip is defined as the difference in the vertex-fit $\chi^2$ of the track's associated PV reconstructed with and without the track under consideration;\footnote{The associated PV is defined as that giving the smallest \chisqip value for the \bquark-hadron candidate.} the $\chi^2$ of the common-vertex fit of the two tracks and the distance of \textcolor{black}{the} closest approach between them.
The BDT also exploits the following properties of the \bquark-hadron candidate: transverse momentum, \chisqip, flight distance (FD), and the $\chi^2$ of the FD, all with respect to the PV associated to the candidate.
The $\chi^2$ of the FD is a measure of how well the decay vertex of the candidate is separated from the PV.
The BDT is trained using Run~2 simulated signal decays and combinatorial background candidates from  Run~2 data in the high-mass sideband, defined as $m_{\pip\pim}>5.6\gevcc$.

Separate optimizations of the selection criteria applied to the BDT output and the PID variables are performed for \LbTopK and \LbToppi decays.
These two selections will be denoted hereafter as \SpK and \Sppi, respectively.
For each selection, the same BDT requirement is used to select the other final-state samples, while different PID requirements are used to separate the \Kp\pim, \pip\Km, \pip\pim, \Kp\Km final states, as well as cross-feed \proton\pim (\proton\Km) and \antiproton\pip (\antiproton\Kp) final states for \SpK (\Sppi).
The optimization procedure has been designed during the Run~1 analysis~\cite{LHCb-PAPER-2018-025} to obtain the best sensitivity on the raw asymmetry from Eq.~\eqref{eq:Araw}, and is based on pseudoexperiments.
The optimization is performed only for the Run~2 sample, while the same BDT and PID selection from the previous analysis is kept for the Run~1 sample~\cite{LHCb-PAPER-2018-025}, \textcolor{black}{given that the additional fiducial requirements in the $(p, \eta)$ plane change the already excellent signal-to-background ratio of the selected candidates only by about 10\%.}

In less then 1\% of the events, more than one candidate is present.
In such cases, all candidates but one are rejected according to a reproducible pseudorandom sequence.

Finally, the dataset is split into 12 disjoint subsets based on the year of acquisition (2011, 2012, 2015, 2016, 2017, 2018) and polarity of the dipole magnet (Up, Down).
This provides two useful checks: firstly, that the measured \CP asymmetries are independent of the data-taking conditions and, secondly, that any possible residual left-right asymmetry has a negligible impact on the determination of $A_{\CP}$.
The \CP asymmetries are measured separately in each subsample and then averaged to obtain the final result of the measurement following the method of Ref.~\cite{Schmelling:1994pz}.

\section{Invariant-mass fit}
\label{sec:InvariantMassFit}
For each final-state hypothesis, namely $\proton \Km$, $\antiproton \Kp$, $\proton \pim$, $\antiproton \pip$, $\Kp\pim$, $\Km\pip$, $\Kp\Km$ and $\pip\pim$, the invariant-mass distribution of selected candidates is modeled by an appropriate probability density function.
These models are used to perform a simultaneous fit to the eight invariant-mass spectra and determine the yields and raw asymmetries of all two-body \mbox{\bquark-hadron} signal decays contributing to the spectra.
Three categories are considered for the background: combinatorial, due to random combinations of oppositely charged particles; partially reconstructed, due to multibody \bquark-hadron decays with one or more particles not reconstructed; and cross-feed, arising from other two-body \bquark-hadron decays where one or both final-state particles are misidentified.

The model used to describe each signal is the sum of a Gaussian function and a Johnson $S_U$ function~\cite{Johnson:1949zj} with common mean and width.
For the high-yield channels, \ie \BdToKpi, \BsToKK, and \BdTopipi, an additional Gaussian function with the same mean and an independent width is added to improve the quality of the fit.
The shape parameters of each model are obtained from fits to simulated samples and fixed in the fit to data, with the exception of the mean and width(s).

The combinatorial background is modeled using exponential functions, with the exception of the \proton\Km and \antiproton\Kp spectra, for which the exponential is multiplied by a function of the form $[1 + \tanh(m)]$, where $m$ is the invariant mass of the hadron pair, to modulate the rise of the exponential at low invariant masses.
This is introduced to model the threshold effect due to the requirements applied in the software trigger on the dipion mass.

The partially reconstructed background is parameterized using ARGUS functions~\cite{ARGUS:1994rms} convolved with Gaussian functions with zero mean and with the same width as used for the signal model in that channel.
For the three signal decays which employ a second Gaussian, the width of the component with the highest fit fraction is used for the convolution.
For the \proton\Km and \proton\pim final states, dedicated templates are built by producing large samples of \decay{\Lb}{\proton\Km\piz} and \decay{\Lb}{\proton\pim\piz} decays using fast simulation~\cite{Cowan:2016tnm}.
\textcolor{black}{Decays of \Lb to final states with heavier particles or additional missing particles are not taken into consideration, as their reconstructed invariant-mass lies outside the fit region.}
The impact of the BDT requirement is modeled by studying its efficiency with fully simulated \LbTopK decays as a function of the cosine of the angle between the flight direction of the reconstructed $b$-hadron and the direction of its reconstructed momentum. This variable is a good proxy for the fraction of momentum removed by the undetected particle in the multibody decays.
The obtained efficiency is then used to correct the samples produced with the fast simulation, where the direction of the \bquark-hadron momentum is obtained considering only the two charged final-state particles.
The produced templates are then used directly in the fit model to describe the shape of partially reconstructed backgrounds.

Finally, the cross-feed backgrounds are modeled by applying a kernel density estimation method~\cite{Cranmer:2000du} to simulated two-body \bquark-hadron decays.
The simulated samples are first weighted using PID efficiency tables as a function of particle kinematics.
The cross-feed background yields are then set to the corresponding two-body \bquark-hadron decay yields, determined by the simultaneous fit, multiplied by appropriate PID-efficiency ratios.
The efficiencies for a given PID requirement are determined from calibration samples in bins of particle momentum, pseudorapidity and hit multiplicity in the SPD, as the performances of the RICH detectors depend on such variables.
They are then averaged over the corresponding distributions of background-subtracted signal candidates.

\begin{figure}[t]
    \includegraphics[width=0.49\textwidth]{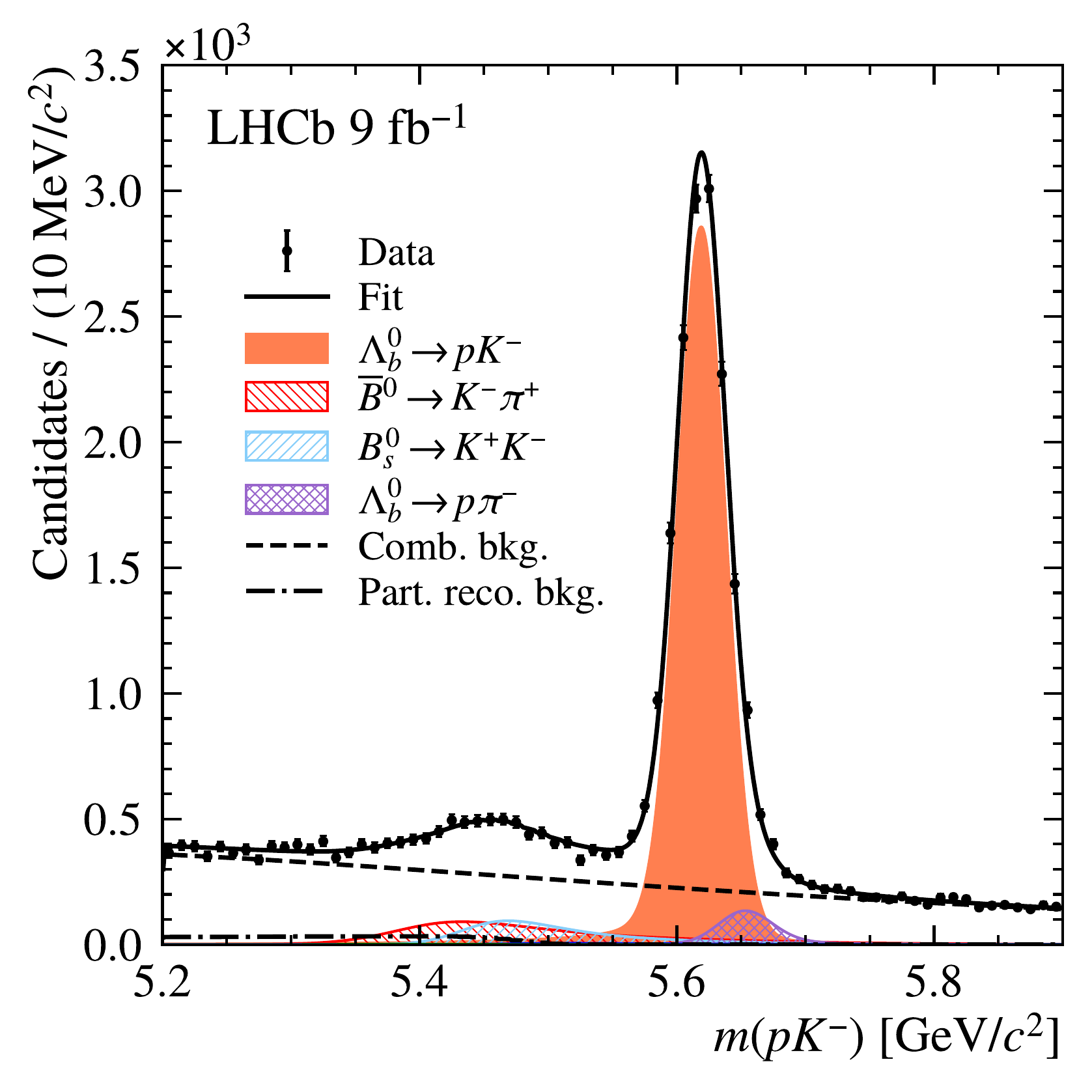}
    \includegraphics[width=0.49\textwidth]{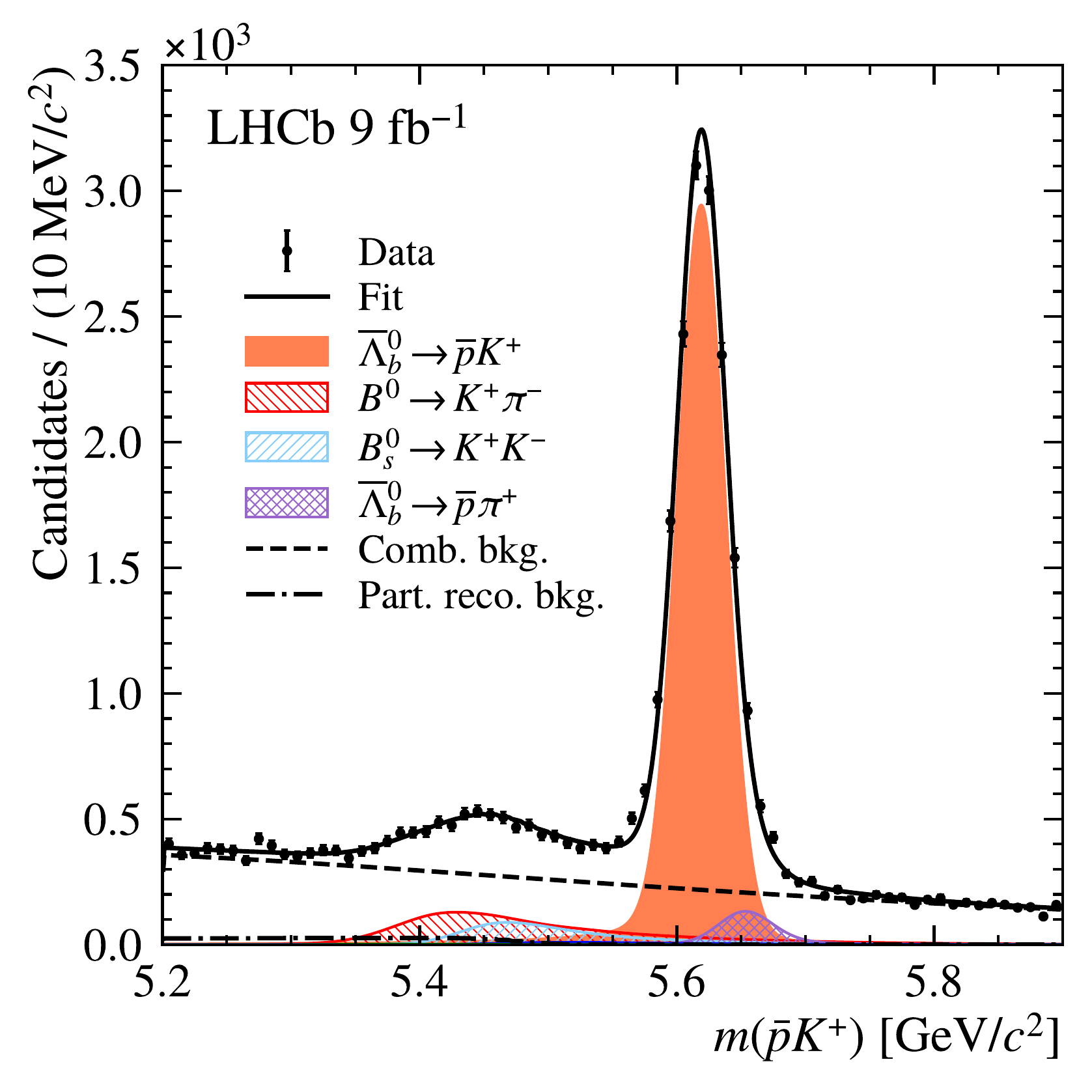}\\
    \includegraphics[width=0.49\textwidth]{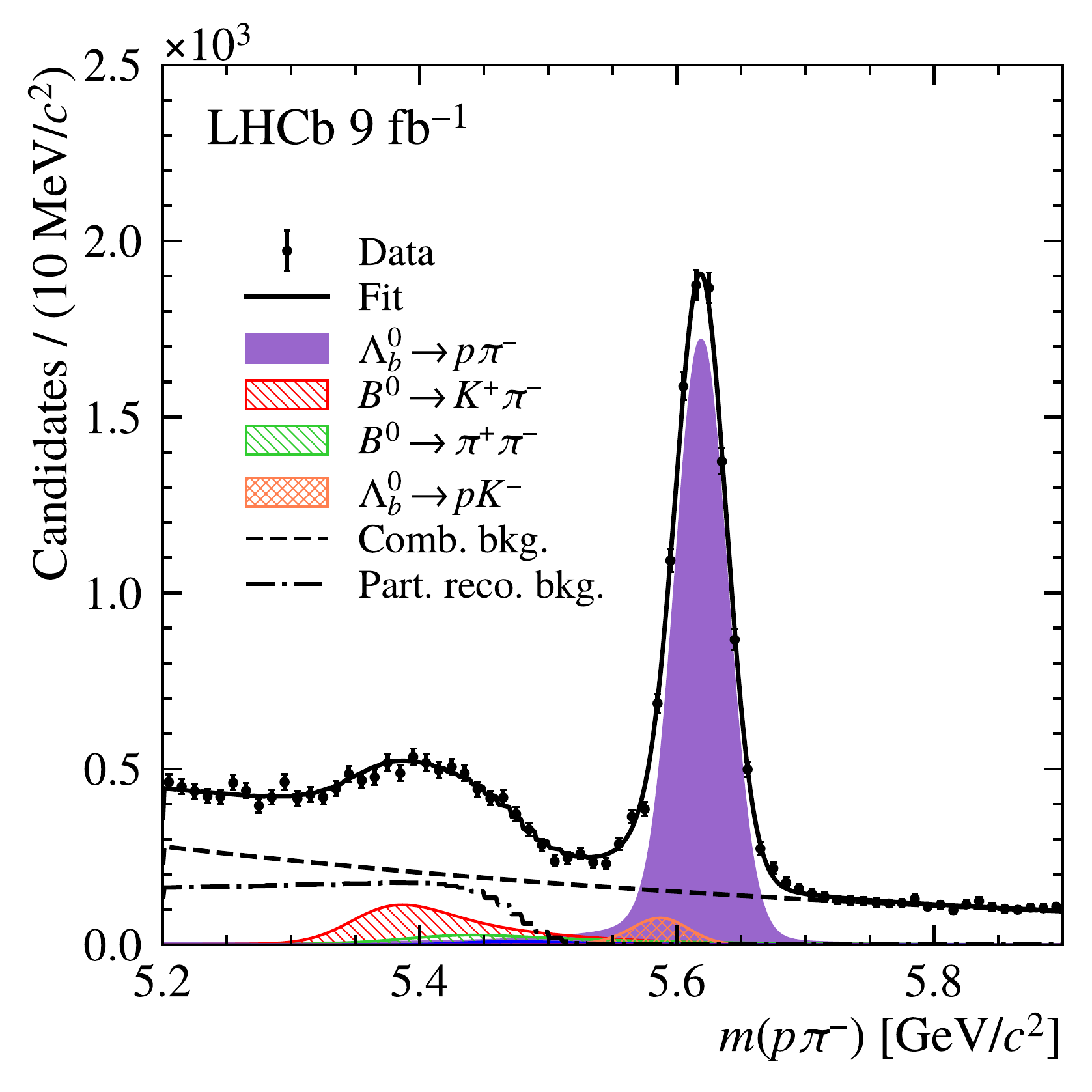}
    \includegraphics[width=0.49\textwidth]{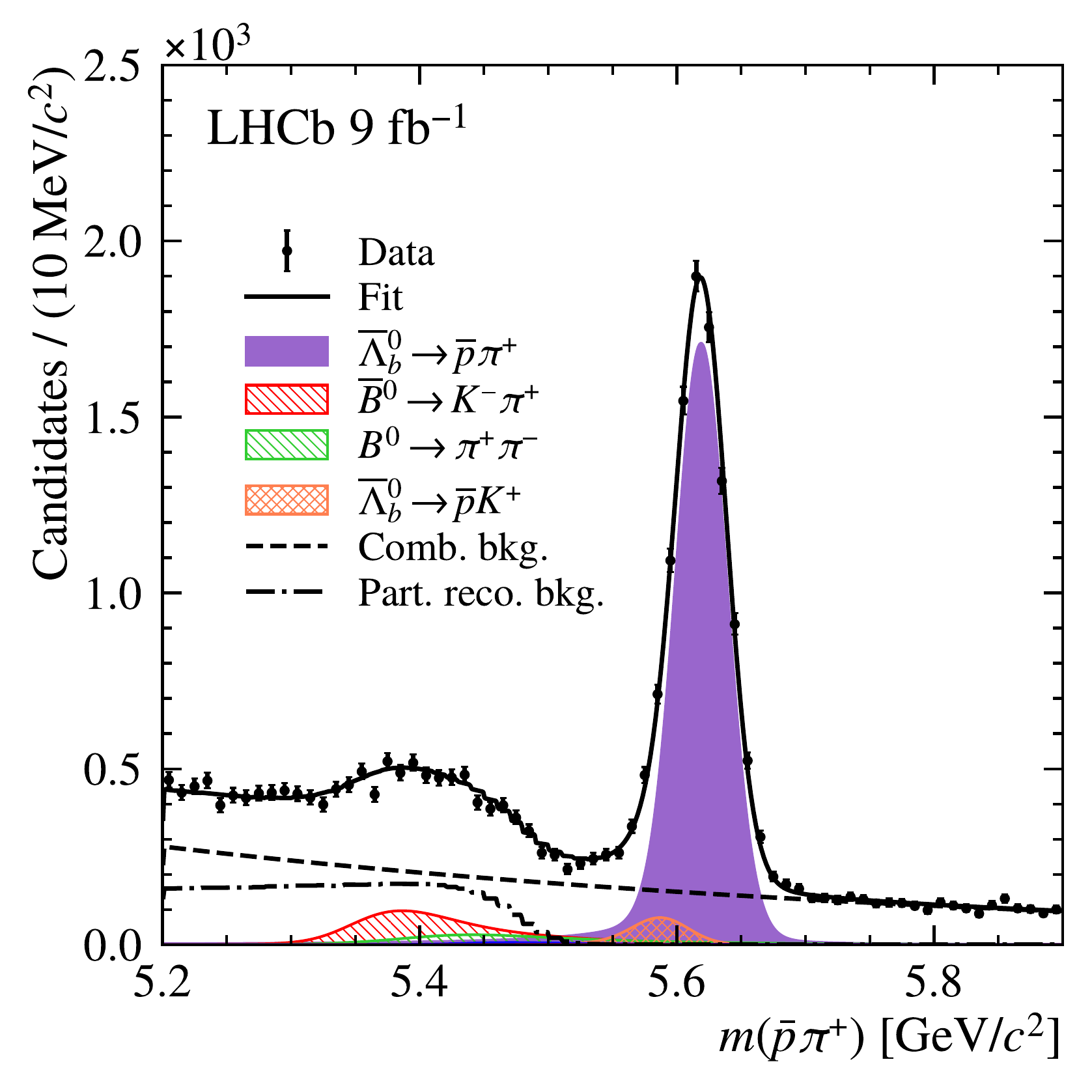}
    \caption{Invariant-mass distributions of the total Run~1 and Run~2 sample: (top left) $m_{p K^-}$, (top right) $m_{\overline{\proton}K^+}$, (bottom left) $m_{p\pi^-}$ and (bottom right) $m_{\overline{p}\pi^+}$ for candidates passing the (top) $S_{pK^-}$ and (bottom) $S_{p\pi^-}$ selections. The results of the fits are also shown.}
    \label{fig:FinalFits}
\end{figure}

After the application of the BDT and PID requirements, an extended binned maximum-likelihood fit is performed simultaneously to the eight final-state samples for each of the two selections, \SpK and \Sppi.
The $m_{\proton \Km}$ and $m_{\proton \pim}$ invariant-mass distributions are shown in Fig.~\ref{fig:FinalFits} after combining all 12 subsamples, with the results of the fits also shown.
\textcolor{black}{About \num{6e3} and \num{4e3} signal events are found in the Run 1 sample for \LbTopK and \LbToppi, respectively, while for Run 2 they are respectively about \num{23e3} and \num{14e3}.}

The fits are validated by generating a large number of pseudoexperiments according to the total probability density function of the model and performing an extended binned maximum-likelihood fit to each sample.
The resulting pull distributions for \ArawpK and \Arawppi are each found to be consistent with a standard normal distribution.

The raw asymmetries in the control sample of \decay{\Lb}{\Lc\pim} decays used in Eqs.~\eqref{eq:ACPpKRun2} and~\eqref{eq:ACPppiRun2} are measured with unbinned maximum-likelihood fits to the invariant mass of the weighted sample split into the same 12 subsets as the signal channel.
The signal is modeled with a Crystal Ball function~\cite{Skwarnicki:1986xj}, the combinatorial category with an exponential function, and the cross-feed background events from \decay{\Lb}{\Lc\Km} with a Gaussian distribution with the same width as the signal function and the mean shifted by the impact of the wrong mass hypothesis.
About \num{168e3} signal events are found for this control sample in Run~1 and Run~2.

\section{Instrumental asymmetries}
\label{sec:InstrumentalAsymmetries}
Determination of the instrumental asymmetries introduced in Eqs.~\eqref{eq:ACPph}, ~\eqref{eq:ACPpKRun2}, and~\eqref{eq:ACPppiRun2} is crucial to obtain the \CP asymmetries, as described in Sec.~\ref{sec:Formalism}. The pion detection asymmetry is obtained by exploiting the measurements in Refs.~\cite{LHCb-PAPER-2016-013} and~\cite{LHCb-PAPER-2021-009} for Run~1 and Run~2, respectively, as a function of pion momentum and pseudorapidity.
The kaon detection asymmetry is determined as a function of the kaon momentum and pseudorapidity, following the approach developed in Ref.~\cite{LHCb-PAPER-2014-013}.
\textcolor{black}{Finally, maps of the proton detection asymmetry, as a function of proton momentum, are taken from Ref.~\cite{LHCB-PAPER-2021-016}. The maps were determined using Run 2 data and applied to both Run 1 and Run 2 samples.}
This approach is validated by the compatibility of the kaon and pion detection asymmetries between Run~1 and Run~2, at the level of $0.6$ standard deviations ($\sigma$) or better.
For the three types of final-state particles, a weighted average of the kinematically dependent asymmetries is evaluated over the background-subtracted~\cite{Pivk:2004ty} kinematic distributions of pions, kaons, and protons from \LbTopK, \LbToppi, and \LbToLcpi decays to obtain the integrated detection asymmetries.

The PID asymmetries are computed according to Eq.~\eqref{eq:PIDAsym}, where the PID efficiencies for a final state $f$ are obtained  by integrating two-dimensional maps of efficiency for a set of PID requirements over the distribution of particle momentum and pseudorapidity taken from simulation.
The decrease in performance with increasing event occupancy is also accounted for. The overall \Lb production asymmetry for Run~1 is calculated by averaging the production asymmetries measured as a function of \Lb rapidity~\cite{LHCB-PAPER-2021-016} over the background-subtracted rapidity distributions of \LbTopK and \LbToppi candidates.

Asymmetries related to different trigger efficiencies for the charge-conjugated final states, \AT{}, may arise.
Signal candidates are classified depending on whether they are directly responsible for the affirmative decision of the trigger (denoted as a \emph{Trigger On Signal}, or TOS candidates) or not (\emph{Trigger Independent of Signal}, or TIS candidates).
In the former case, an asymmetry can arise due to different probabilities of positively or negatively charged hadrons firing the trigger; in the latter, the same can happen to the decay products of the other \bquark-hadron produced in the \proton\proton collision.
For candidates classified as TOS at both the hardware and software stages of the \lhcb trigger, the corresponding efficiency is determined using samples of \decay{\Lb}{(\decay{\Lc}{\proton\Km\pip})\mu^-\nu X} for protons and \decay{\Bz}{(\decay{\Dzb}{\Kp\pim})\mu^+\nu X} decays for pions and kaons, where only the muon is required to give a positive trigger decision.
For the hardware trigger, asymmetries are determined as a function of the transverse-energy deposit in the calorimeter system.
For the software trigger, the \pt and $\log(\chisqip)$ of the decay products are considered, as these are the variables used by the corresponding algorithms to select or reject candidates.
Systematic uncertainties for this procedure are evaluated by applying the same method on simulated samples and comparing the results with the asymmetries obtained using truth information.
For TIS candidates, which instead are only selected by the hardware trigger stage, the charge asymmetry is determined by studying a sample of \decay{\Bp}{\jpsi\Kp} decays~\cite{LHCb-DP-2019-001} and computing the TIS efficiency as a function of \Bp transverse momentum.
The TIS asymmetry is then obtained by averaging the \pt-dependent values over the background-subtracted kinematic distributions of \LbTopK and \LbToppi decays.

\section{Systematic uncertainties and combination}
\label{sec:SystematicUncertaintiesAndCombination}
Several sources of systematic uncertainties associated with the fit model are investigated.
Alternative models are considered: the baseline model used for signal is replaced by a double-sided Crystal Ball function~\cite{Skwarnicki:1986xj}; the templates for the cross-feed decays are modified by excluding the correction for the PID efficiencies; the function used to model the combinatorial background is substituted with a second-order polynomial; and the templates describing the partially reconstructed backgrounds in the \proton\Km, \antiproton\Kp, \proton\pim, and \antiproton\pip spectra, defined in Sec.~\ref{sec:InvariantMassFit}, are replaced by ARGUS functions~\cite{ARGUS:1994rms}.
When testing alternative models, 1000 pseudoexperiments are generated according to the baseline fit results.
Fits are performed to each of the generated samples using the baseline model and then the alternative models.
The mean and the root mean square of the distribution of the difference between the raw asymmetries determined by the two sets of fits are added in quadrature with the resulting value taken as a systematic uncertainty.

A different approach is adopted to assess systematic uncertainties related to the knowledge of the cross-feed background contamination.
Samples are generated using the baseline fit results.
The baseline model is then fitted 1000 times to the generated samples, varying the PID efficiencies according to their uncertainties, which are mainly driven by the choice of the binning scheme used to divide the phase space.
On changing the binning scheme and recomputing the efficiencies, the largest deviation from the baseline is taken as a systematic uncertainty.
Finally, the statistical uncertainty on the raw asymmetry extracted from the \LbToLcpi control sample is propagated to the $A_{\CP}$ measurement, and quoted as a systematic uncertainty.

Correlations between the results in the various subsamples may arise from the use of calibration samples that are common among them, \eg the use of the proton detection asymmetry measurement from Run 1 over all the Run 2 samples, or the computation of several instrumental asymmetries on the signal and control samples following Eqs.~\eqref{eq:ACPpKRun2} and~\eqref{eq:ACPppiRun2} with the same method.
These correlations are evaluated by generating 1000 versions of the tables used to compute all the corrections, sampled from Gaussian distributions with means equal to the central values and widths equal to the nominal uncertainties.
Then the 1000 sets of corrections are used to recompute \ACPpK and \ACPppi.
From the distributions of the 1000 values the correlations between the various subsamples are inferred.
The correlations are found to be smaller than 8\%, but are taken into account nonetheless when averaging the results.

To quantify the size of each systematic uncertainty on the final measurement, the bootstrapping procedure is repeated many times, each one varying only the baseline tables corresponding to a single correction, to decouple them from the others to which they might be correlated.
This allows a breakdown of the contribution of each systematic uncertainty to the final results, which is reported in~\Cref{tab:systematic}.
Some systematic uncertainties are applicable to either Run~1 or Run~2 asymmetries.
For instance, the use of the \LbToLcpi control sample for Run~2 requires the evaluation of detection asymmetries for protons, kaons, and pions, as opposed to the Run~1 sample in which only two of these particles are needed for each signal channel.
As this pertains to the \Lb production asymmetry, its contribution for Run~2 is removed by the kinematic weighting of the \LbToLcpi sample, as explained in Eqs.~\eqref{eq:ACPpKRun2} and~\eqref{eq:ACPppiRun2}.
Conversely, this control sample is not needed for Run~1 as the relevant \Lb production asymmetry is measured by other means.
\begin{table}
    \caption{Absolute systematic and statistical uncertainties, in percent, on the \CP asymmetry measurements, split by decay channel and data-taking period. The dashes represent systematic uncertainties that do not contribute to the measurement in the corresponding sample.}
    \label{tab:systematic}
    \centering
    \begin{tabular}{rcccc}
        \toprule
                                 & \multicolumn{2}{c}{Run 1}      & \multicolumn{2}{c}{Run 2}                                                                      \\
        \cmidrule(r){2-3} \cmidrule(r){4-5}
                                 & \LbTopK                        & \LbToppi                      & \LbTopK                        & \LbToppi                      \\
        \midrule
        Fit model                & \rounded{0.05}                 & \rounded{0.15}                & \rounded{0.05}                 & \rounded{0.15}                \\
        Particle identification  & \rounded{0.25}                 & \rounded{0.25}                & \rounded{0.15}                 & \rounded{0.16}                \\
        TIS trigger              & \rounded{0.12}                 & \rounded{0.11}                & $<0.1$                         & $<0.1$                        \\
        TOS hardware trigger     & \rounded{0.20}                 & \rounded{0.21}                & \rounded{0.10}                 & \rounded{0.10}                \\
        TOS software trigger     & \rounded{0.33}                 & \rounded{0.32}                & \rounded{0.20}                 & \rounded{0.20}                \\
        Proton detection         & \rounded{0.10}                 & \rounded{0.10}                & $<0.1$                         & $<0.1$                        \\
        Kaon detection           & \rounded{0.25}                 & --                            & \rounded{0.10}                 & $<0.1$                        \\
        Pion detection           & --                             & \rounded{0.10}                & $<0.1$                         & $<0.1$                        \\
        $\Lambda_b^0$ production & \rounded{0.12}                 & \rounded{0.13}                & --                             & --                            \\
        \LbToLcpi sample size    & --                             & --                            & \rounded{0.28}                 & \rounded{0.28}                \\
        \midrule
        Total systematic         & $ \rounded{\ACPpkRunOneSyst}
        $                        & $ \rounded{\ACPppiRunOneSyst}$ & $ \rounded{\ACPpkRunTwoSyst}$ & $ \rounded{\ACPppiRunTwoSyst}$                                 \\
        Statistical              & $\rounded{\ACPpkRunOneStat}$   & $\rounded{\ACPppiRunOneStat}$ & $\rounded{\ACPpkRunTwoStat}$   & $\rounded{\ACPppiRunTwoStat}$ \\
        \bottomrule
    \end{tabular}
\end{table}

\section{Results and conclusions}
\label{sec:ResultsAndConclusions}
The measurements of the \CP-violating asymmetries in the 12 individual data-taking year and magnet-polarity subsamples are found to be consistent, with a $\chi^2/\mathrm{ndf}$ of 11.6/11 for \ACPpK and 6.7/11 for \ACPppi, showing no dependence on the data-taking conditions.
Using Eq.~\eqref{eq:ACPph} in each subsample and after computing their average value according to Ref.~\cite{Schmelling:1994pz}, the Run~1 \CP asymmetries are measured to be
\begin{align*}
    \ACPpK  & = (\rounded{\ACPpkRunOneCentral} \pm \rounded{\ACPpkRunOneStat} \pm \rounded{\ACPpkRunOneSyst})\%,    \\
    \ACPppi & = (\rounded{\ACPppiRunOneCentral} \pm \rounded{\ACPppiRunOneStat} \pm \rounded{\ACPppiRunOneSyst})\%,
\end{align*}
where the first uncertainties are statistical and the second are systematic, with the correlation between the two measurements being $\ACPcorrRunOne\%$.
The results are compatible with the previous result from LHCb using the same dataset~\cite{LHCb-PAPER-2018-025}.
While the updated measurements have larger statistical uncertainties due to the PID fiducial requirements, the systematic and total uncertainties are significantly reduced, in particular due to the new determination of the \Lb production asymmetry~\cite{LHCB-PAPER-2021-016}.
As an additional cross-check, the method formalized in Eqs.~\eqref{eq:ACPpKRun2} and~\eqref{eq:ACPppiRun2} is also used to measure the Run~1 \CP asymmetries, obtaining compatible results.
The Run~2 \CP asymmetries, measured following Eqs.~\eqref{eq:ACPpKRun2} and~\eqref{eq:ACPppiRun2}, are similarly averaged over the Run~2 samples to obtain
\begin{align*}
    \ACPpK  & = (\rounded{\ACPpkRunTwoCentral} \pm \rounded{\ACPpkRunTwoStat} \pm \rounded{\ACPpkRunTwoSyst})\%,               \\
    \ACPppi & = (\phantom{-}\rounded{\ACPppiRunTwoCentral} \pm \rounded{\ACPppiRunTwoStat} \pm \rounded{\ACPppiRunTwoSyst})\%,
\end{align*}
with a correlation of $\ACPcorrRunTwo\%$.
These values are compatible with the Run~1 results, with which they are combined to obtain the final results
\begin{align*}
    \ACPpK  & = (\rounded{\ACPpkCentral} \pm \rounded{\ACPpkStat} \pm \rounded{\ACPpkSyst})\%,               \\
    \ACPppi & = (\phantom{-}\rounded{\ACPppiCentral} \pm \rounded{\ACPppiStat} \pm \rounded{\ACPppiSyst})\%,
\end{align*}
with a total correlation of $\ACPcorrRunOneTwo\%$. By performing a scan, the likelihood of both \CP asymmetries has been found to be Gaussian.
No evidence of \CP violation is found.
These are the most precise measurements of these observables to date, improving over the current world averages by a factor of three.
These results supersede those previously obtained by LHCb~\cite{LHCb-PAPER-2018-025}, and will help in constraining theoretical models for QCD effects contributing to these two-body baryonic decays.

\section*{Acknowledgements}
%
%
\noindent We express our gratitude to our colleagues in the CERN
accelerator departments for the excellent performance of the LHC. We
thank the technical and administrative staff at the LHCb
institutes.
We acknowledge support from CERN and from the national agencies:
CAPES, CNPq, FAPERJ and FINEP (Brazil);
MOST and NSFC (China);
CNRS/IN2P3 (France);
BMBF, DFG and MPG (Germany);
INFN (Italy);
NWO (Netherlands);
MNiSW and NCN (Poland);
MCID/IFA (Romania);
MICIU and AEI (Spain);
SNSF and SER (Switzerland);
NASU (Ukraine);
STFC (United Kingdom);
DOE NP and NSF (USA).
We acknowledge the computing resources that are provided by CERN, IN2P3
(France), KIT and DESY (Germany), INFN (Italy), SURF (Netherlands),
PIC (Spain), GridPP (United Kingdom),
CSCS (Switzerland), IFIN-HH (Romania), CBPF (Brazil),
and Polish WLCG (Poland).
We are indebted to the communities behind the multiple open-source
software packages on which we depend.
Individual groups or members have received support from
ARC and ARDC (Australia);
Key Research Program of Frontier Sciences of CAS, CAS PIFI, CAS CCEPP,
Fundamental Research Funds for the Central Universities,
and Sci. \& Tech. Program of Guangzhou (China);
Minciencias (Colombia);
EPLANET, Marie Sk\l{}odowska-Curie Actions, ERC and NextGenerationEU (European Union);
A*MIDEX, ANR, IPhU and Labex P2IO, and R\'{e}gion Auvergne-Rh\^{o}ne-Alpes (France);
AvH Foundation (Germany);
ICSC (Italy);
Severo Ochoa and Mar\'ia de Maeztu Units of Excellence, GVA, XuntaGal, GENCAT, InTalent-Inditex and Prog. ~Atracci\'on Talento CM (Spain);
SRC (Sweden);
the Leverhulme Trust, the Royal Society
and UKRI (United Kingdom).

\newpage
\addcontentsline{toc}{section}{References}
\setboolean{inbibliography}{true}
\ifx\mcitethebibliography\mciteundefinedmacro
    \PackageError{LHCb.bst}{mciteplus.sty has not been loaded}
    {This bibstyle requires the use of the mciteplus package.}\fi
\providecommand{\href}[2]{#2}

\newpage
\centerline
{\large\bf LHCb collaboration}
\begin{flushleft}
\small
R.~Aaij$^{38}$\lhcborcid{0000-0003-0533-1952},
A.S.W.~Abdelmotteleb$^{57}$\lhcborcid{0000-0001-7905-0542},
C.~Abellan~Beteta$^{51}$\lhcborcid{0009-0009-0869-6798},
F.~Abudin{\'e}n$^{57}$\lhcborcid{0000-0002-6737-3528},
T.~Ackernley$^{61}$\lhcborcid{0000-0002-5951-3498},
A. A. ~Adefisoye$^{69}$\lhcborcid{0000-0003-2448-1550},
B.~Adeva$^{47}$\lhcborcid{0000-0001-9756-3712},
M.~Adinolfi$^{55}$\lhcborcid{0000-0002-1326-1264},
P.~Adlarson$^{82}$\lhcborcid{0000-0001-6280-3851},
C.~Agapopoulou$^{14}$\lhcborcid{0000-0002-2368-0147}
% ...

{\footnotesize \it

% ...
$^{14}$Universit{\'e} Paris-Saclay, CNRS/IN2P3, IJCLab, Orsay, France\\
% ...
$^{38}$Nikhef National Institute for Subatomic Physics, Amsterdam, Netherlands\\
% ...
$^{47}$Instituto Galego de F{\'\i}sica de Altas Enerx{\'\i}as (IGFAE), Universidade de Santiago de Compostela, Santiago de Compostela, Spain\\
% ...
$^{51}$Physik-Institut, Universit{\"a}t Z{\"u}rich, Z{\"u}rich, Switzerland\\
% ...
$^{55}$H.H. Wills Physics Laboratory, University of Bristol, Bristol, United Kingdom\\
% ...
$^{57}$Department of Physics, University of Warwick, Coventry, United Kingdom\\
% ...
$^{60}$School of Physics and Astronomy, University of Glasgow, Glasgow, United Kingdom\\
$^{61}$Oliver Lodge Laboratory, University of Liverpool, Liverpool, United Kingdom\\
% ...
$^{69}$Syracuse University, Syracuse, NY, United States\\
% ...
$^{82}$Department of Physics and Astronomy, Uppsala University, Uppsala, Sweden, associated to $^{60}$\\
% ...
}
\end{flushleft}
\end{document}